\documentstyle[multicol,prl,aps,epsf]{revtex}
\begin{document}
\setlength{\topmargin}{-0.6in}

\title{Can half metallic zincblende MnAs be grown?}
\author{Stefano Sanvito\thanks{e-mail: ssanvito@mrl.ucsb.edu} and Nicola A. Hill}
\address{Materials Department, University of California,
Santa Barbara, CA 93106, USA}
\date{\today}
\maketitle

\begin{abstract}
In this paper we investigate theoretically the ground state
NiAs-type structure of MnAs and we compare the
magnetic and structural properties with an hypothetical zincblende structure.
A zincblende structure can be obtained, in principle, from the
diluted magnetic semiconductor Ga$_{1-x}$Mn$_x$As in the limit $x=1$.
Using density functional calculations within the local spin-density 
approximation (LSDA), we show that 
the zincblende structure, although showing half metallic behavior which is 
very attractive for spintronics, can not be stabilized at equilibrium.
We perform a tight-binding analysis of the Mn-As bond in the tetrahedral 
coordination to investigate the nature of the bonding in
diluted magnetic semiconductors.
\end{abstract}

\begin{multicols}{2}

\vspace{0.3in}


\narrowtext

\section{Introduction}
In the last few years there has been a growing interest in combining
magnetotransport experiments based onto the giant 
magnetoresistance (GMR) in metallic magnetic multilayers with semiconductor
physics \cite{Pri95}. 
Semiconductors offer various advantages over metals such as a very 
long spin-lifetime \cite{Aws1}, persistent spin coherence \cite{Aws2}, 
and compatibility with existing semiconductor processing technologies.
These properties pave the way for using semiconductors as media for storing 
coherence and realizing elements for solid-state quantum computation \cite{QC}. 
The natural extension of GMR to semiconductors is in a hybrid spin-valve where
the magnetic elements are either magnetic metals or diluted magnetic
semiconductors \cite{Ohno99,Ohno98} and the non-magnetic elements are 
ordinary semiconductors. In both cases it is essential to inject spins into 
the non-magnetic semiconductors and this is a formidable challenge.

So far spin-injection into semiconductors from metallic contacts has been 
elusive \cite{Rouk00} with GMR signals smaller than 1\% \cite{Lee99,Ham99}. 
In contrast spin-injection from diluted magnetic semiconductors has been 
more successful and evidence of polarized currents either into III-V \cite{Awch99} 
and II-VI \cite{Fied99} semiconductors has been demonstrated. In
both cases the polarization is observed by optical techniques, which
makes the integration of such systems in present semiconductor technology very
difficult and all electronic systems would be more technologically desirable.
Recently it has been suggested \cite{Schm99} that a spin-valve with a conventional
injector-detector geometry made by metallic contacts cannot present large
signals at equilibrium due to the very different resistivities of the contacts 
and the semiconductor spacer.
Much larger signals are predicted by using half-metallic
contacts or diluted magnetic semiconductors.

From the point of view of the materials the system formed by GaAs as non-magnetic 
semiconductor and by MnAs and Ga$_{1-x}$Mn$_x$As respectively as magnetic metal 
and diluted magnetic semiconductor is the most promising for spintronics 
applications.
For the metallic component in metal/semiconductor structures, 
MnAs presents several advantages over transition metals. 
MnAs grows epitaxially onto GaAs in the hexagonal NiAs-type structure. 
This structure is characteristic of all bulk MnX compounds with X=As, Sb and Bi.
High quality all-epitaxial MnAs structures can be grown onto commonly used GaAs 
\cite{Tan99-1,Tan93,Tan94-1} and Si \cite{Ake95}.
For MnAs/GaAs the interfaces are thermodynamically stable since the
materials share the As atoms and the growth process is completely 
compatible with existing III-V MBE technology. 

Another attractive prospective crystal structure for MnAs is the zincblende structure. 
Recent density functional calculations \cite{Oga99} have shown that the 
latter, which can be obtained in principle in the limit of 100\% 
doping from Ga$_x$Mn$_{1-x}$As, is a half-metal.
This makes zincblende MnAs very promising for spintronics because it will 
allow the intrinsic difficulty of injecting spins into semiconductors 
to be overcome \cite{Schm99}. 
Unfortunately to our knowledge zincblende MnAs has never been successfully 
grown.

In this paper we investigate, by using density functional calculations, the
stability of the NiAs-type MnAs with respect to the hypothetical zincblende
structure. In particular we look for growth conditions (for example strains 
from the growing substrate, volume compressions) which 
can enable the zincblende phase to be made. 
One of the main results of this analysis is that the zincblende structure 
cannot be stabilized either by lattice stretching nor by compression. 
We also consider the magnetic properties of the NiAs-type MnAs grown onto 
GaAs with different crystalline orientations and lattice distortions. 
Finally we look at the nature of the chemical bonding in zincblende MnAs. This is
particularly important since the tetrahedral coordination of the Mn atoms is one
of the key elements for understanding the properties of magnetic diluted III-V
semiconductors.

The remainder of this paper is organized as follows. 
In the next sections we will briefly discuss the calculation technique and provide 
some information regarding the lattice structures considered and their properties. 
Then we will move on to analyze the stability of the
NiAs-type MnAs under compression and distortion of the unit cell. In
section V we will consider the zincblende structure and we will compare its
structural properties with the NiAs-type. Finally we will 
perform a fit onto a tight-binding model in order to analyze the details of
the chemical bonding in zincblende MnAs.

\section{Computational Details}

We compute the electronic structure of MnAs with both NiAs-type and zincblende
lattice structures using a plane wave pseudopotential implementation of
density functional theory \cite{Kohn64} with the local spin density
approximation. The use of pseudopotentials for the study of magnetic systems, although 
not yet as popular as all electron methods,
is now well documented \cite{Nicola99}. We use the optimized
pseudopotentials developed by Rappe {\it et al.} \cite{Rap90}
for our Mn pseudopotential, allowing us to use a 
small cutoff energy for the plane wave expansion in the total energy 
calculation of the solid. This is achieved by minimizing the kinetic energy in
the high Fourier components of the pseudo wave function.
The reference configuration for our Mn is 4s$^{0.75}$4p$^{0.25}$3d$^5$ with 
core radii of 2.0, 2.15 and 2.0~a.u. and we use the usual Kleinman-Bylander separable
form with two projectors for each angular momentum \cite{Kle82,Blo90}.
For As we use the standard Hamann-Sch\"ulter-Chiang pseudopotential \cite{Ham79} 
with one projector for each angular momentum.
The pseudopotentials were tested by comparing the pseudoeigenvalues with those
generated by all electron calculations for several atomic and ionic
configurations. We also checked the portability by calculating the equilibrium
lattice constant of GaAs and the dependence of the magnetization on the unit
cell volume of hypothetical fcc Mn. Both are in excellent agreement with all
electron calculations. Moreover the Mn pseudopotentials has been successfully used in
earlier studies of perovskite manganites \cite{Nicola99}.

Total energy and band structure are calculated by using the code
SPECTER \cite{Ger00}. This is a spin-polarized implementation of density functional theory
written in C, which originated from the program CASTEP 2.1 \cite{Payne92}.
The plane wave cutoff is fixed to 870~eV ($\sim$63~Ry) and we use
a simple electron density mixing scheme for convergence developed by
Kerker \cite{Kerk81}
with variable Gaussian broadening between 0.8 and 10$^{-4}$ eV.
We use a $4\times 4 \times 4$ Monkhorst-Pack grid for the zincblende
structure and a $3\times 3 \times 3$ for the NiAs-type structure. This leads
respectively to 10 and 6 $k$-points in the corresponding irreducible Brillouin 
zones. Stability of the results with respect to $k$-point sampling has been
carefully checked and the present choice represents a good compromise between
accuracy and computation time.

\section{Crystal Structures}

\begin{figure}[htbp]
\narrowtext
\epsfysize=6cm
\epsfxsize=9cm
\centerline{\epsffile{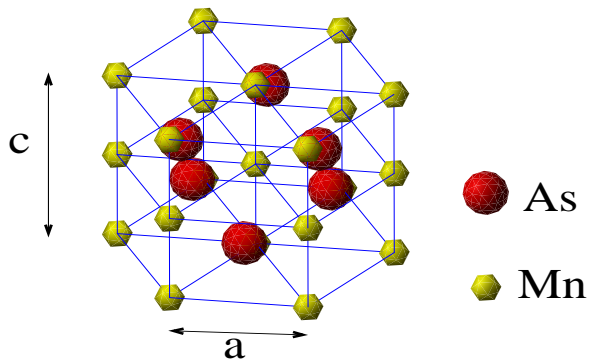}}
\caption{NiAs-type crystal structure. The lattice constants for bulk MnAs
are a=3.7\AA\ and c=5.7\AA\ .}
\label{fig1}
\end{figure}

The NiAs-type lattice structure is an hexagonal structure (space group P$6_3$/mmc)
with 4 atoms in the primitive unit cell (see figure \ref{fig1}). 
It is the equilibrium lattice structure of the pnictides formed with Mn at room 
temperature. Bulk MnAs has lattice constants of a=3.7\AA\ and c=5.7\AA\ and is
a ferromagnetic metal with a T$_c$ of 318K. Above T$_c$
it undergoes a phase transition to a paramagnetic state involving 
also a structural change to the hexagonal MnP-type lattice structure 
(space group Pnma) with a 2\% volume reduction 
\cite{Moti86}. A further structural phase transition 
back to the NiAs-type phase is observed at 398K with the material
remaining paramagnetic.
MnAs can grow epitaxially onto GaAs in several possible crystalline
orientations with respect to the GaAs substrate, depending on the growth 
conditions. In particular when MnAs grows onto (001) GaAs two orientations
can occur \cite{Tan94}. The first one, which is usually called type A, has growth
plane ($\bar{1}$100) and epitaxial relationship 
[$\bar{1}\bar{1}$20]MnAs$\parallel$[110]GaAs and [0001]MnAs$\parallel$[$\bar{1}$10]GaAs, while the
second, known as type B, has growth planes ($\bar{1}$101) and
($\bar{1}$102) and epitaxial relationship
[$\bar{1}\bar{1}$20]MnAs$\parallel$[$\bar{1}$10]GaAs and 
[1$\bar{1}$20]MnAs$\parallel$[110]GaAs.
Moreover MnAs can be grown onto (111) GaAs \cite{Tan99}
with growth direction along the c
axis of the hexagonal cell and epitaxial relationship (0001)MnAs$\parallel$(111)GaAs,
[$\bar{1}$100]MnAs$\parallel$[11$\bar{2}$]GaAs and 
[$\bar{2}$110]MnAs$\parallel$[1$\bar{1}$0]GaAs.
All these phases have the easy axis of magnetization
along the c axis of the hexagonal cell.
The saturation magnetizations (M$_s$) are very different, being
416~emu/cm$^3$, 331~emu/cm$^3$ \cite{Tan94} and 640~emu/cm$^3$ \cite{Tan99}
respectively for type A, type B and (111) GaAs grown MnAs. 
Moreover for MnAs MBE-deposited onto GaAs the saturation magnetization depends on
the number of monolayers deposited \cite{Tan94b}, suggesting that relaxation of 
the lattice has an important influence on the magnetic properties.
In what follows we
will show that these changes of the saturation magnetization can be correlated
with distortions of the hexagonal unit cell resulting from matching to the substrate.

Bulk zincblende MnAs is unstable, and zincblende structure thin films have not been 
demonstrated conclusively \cite{Sam}. However 
Mn can be incorporated into epitaxial layers of GaAs with concentrations beyond 
its solubility limit by using low temperature MBE techniques \cite{Ohno99}. 
The highest concentration obtained so far is $x=0.07$ for Ga$_{1-x}$Mn$_x$As,
above which the Mn atoms segregate forming clusters of MnAs with NiAs-type structure.
Nevertheless the low concentration limit is reproducible. Mn in GaAs acts both
as a source of localized spin and also as an acceptor, providing electrical
carriers (holes). (Ga,Mn)As is ferromagnetic for concentration above $x=0.005$
although the origin of the ferromagnetism is still a matter of debate. In
section V we will study the limit of $x=1$ in order to understand the nature of
the chemical bond of Mn tetrahedrally coordinated with As. 

\section{NiAs-type MnAs}

\begin{figure}[htbp]
\narrowtext
\epsfysize=7.0cm
\epsfxsize=9cm
\centerline{\epsffile{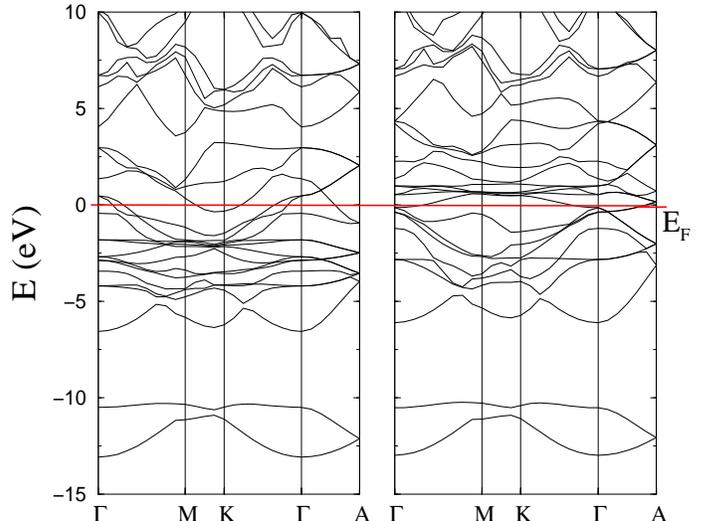}}
\caption{Band structure for bulk MnAs with NiAs-type lattice structure. The figure
on the left corresponds to the majority spin and the one to the right to the
minority. The horizontal line denotes the position of the Fermi energy, which has been chosen
to be 0~eV.}
\label{Fig2}
\end{figure}

\begin{figure}[htbp]
\narrowtext
\epsfysize=6.5cm
\epsfxsize=9cm
\centerline{\epsffile{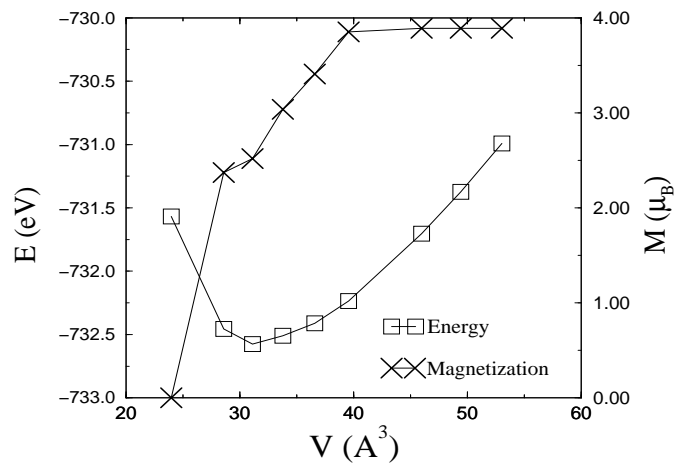}}
\caption{Total energy ($\Box $ on the left-hand side scale) and magnetization 
($\times$ on the right-hand side scale) as a function of
the volume occupied by a MnAs pair.}
\label{Fig3}
\end{figure}

\begin{figure}[htbp]
\narrowtext
\epsfysize=7.0cm
\epsfxsize=9cm
\centerline{\epsffile{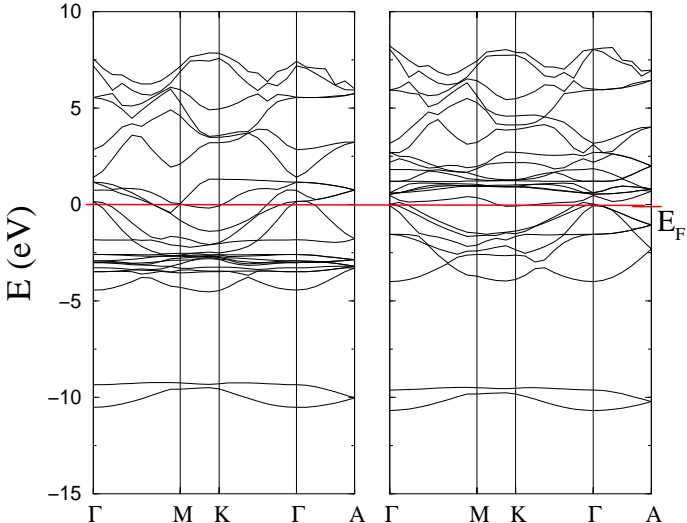}}
\caption{Band structure of NiAs-type MnAs at expanded volume.
The lattice constants are a=4.3\AA\ and c=6.62\AA\ .
The figure
on the left shows the majority spin and the one to the right the
minority. The horizontal line denotes the position of the Fermi energy, 
which has been chosen to be 0~eV.}
\label{Fig4}
\end{figure}

In this section we look at the magnetic
properties and structural stability of the NiAs-type phase.
In figure \ref{Fig2} we present the band structure for ferromagnetic MnAs with lattice
parameter corresponding to the experimental bulk values (a=3.7\AA\ c=5.7\AA\ ). 
The two lower bands correspond to the 4s states of As and the following 16 to
hybridized p-d bands. This band structure is characteristic of all the Mn-based
pnictides \cite{Moti86} and we can identify three distinct regions: i) a low
energy p-d bonding region (between -6 and -2.5~eV for the majority band and
between -6 and 0~eV for the minority), ii) an almost dispersionless intermediate
region ($\sim$-2.0 for the majority and $\sim 1$ for the majority band) and iii)
a high energy p-d antibonding region (between -1.5 and 3~eV for the majority band and
between 1.5 and 4~eV for the minority). The ferromagnetism is mostly due to the
spin-splitting of the degenerate d bands, although a rigid shift model cannot be applied
due to the strong p-d interaction. We will see in the next section that p-d interaction
is also very strong in the zincblende structure.
Turning our attention to the structural properties we calculate the total
energy and the magnetic moment per MnAs pair as a function of the volume
occupied by a pair of atoms. We fix the ratio between the hexagonal axes to 
c/a=1.54, which corresponds to the experimental value for bulk MnAs. The results
are presented in figure \ref{Fig3}.
Our calculated equilibrium volume is 31.123~\AA$^3$ which is slightly smaller
than the experimental one of 33.789~\AA$^3$. This small discrepancy can be
explained by a general tendency of the LDA
to underestimate the equilibrium volume. We also note that the structure is very
stable with respect to compression and expansion of the unit cell around the
equilibrium value (volume changes of the order of 50\% give energy changes of
approximately 1~eV).
Our calculated magnetization $M_s$ at equilibrium is between 2.5 and 
$3.0\mu_B$ which is smaller than that found
experimentally ($\sim 3.4\mu_B$). Nevertheless it is notable that 
the magnetization is very sensitive to the volume around the
equilibrium position, and small volume changes can produce large changes in the
magnetization. 
For large compressions the system evolves to a paramagnetic state. This
transition is not directly comparable with experiments on MnAs under pressure
\cite{Maki98} since a structural transition to the MnP-type structure and a
magnetic transition to a mixed ferromagnetic and antiferromagnetic phase are
experimentally observed for pressures above 4~kbar.
In contrast, for volumes larger than the equilibrium volume we observe a
saturation of the magnetic moment per MnAs pair to a value around 
4$\mu_B$. This saturation can be understood by comparing the band
structures of Fig.\ref{Fig2}
with those obtained for an expanded structure (a=4.3\AA\, c=6.62\AA\
with a volume per MnAs pair of 53.03\AA$^3$) presented in Fig.\ref{Fig4}. 
First we observe 
that there is a large shrinking of the Mn d bands without a drastic change of 
their centers with respect to the Fermi energy. As a result of the strong p-d interaction
the minority p bands below the Fermi energy are pushed to higher energies,
while the majority p bands above the Fermi energy are pulled to lower energy.
This increases the number of occupied majority states and decreases the occupation 
of minority states resulting in a global increase in the magnetic moment 
of the system.
The saturation of the magnetization to a value of $\sim4\mu_B$
can be explained by considering charge transfer from Mn to As. Three electrons will
be transferred from each Mn to a neighboring As in order to fill completely the 
electronegative As atoms' 4p shell.
This leads to an atomic configuration of Mn with four electrons in the
valence, which arrange themself following Hund's rules and produce a magnetic
moment of 4$\mu_B$. The charge transfer is not complete for unit cell
volume corresponding to the experimental value because of the strong p-d interaction.
This p-d interaction 
in fact leads also to a quite large negative polarization of the anions
(-0.23$\pm$0.05$\mu_B$),
which has been observed with neutron scattering \cite{Yam83}.
For stretched unit cells, the p-d hybridation is reduced and a more ionic
configuration is resumed. 

\begin{figure}[htbp]
\narrowtext
\epsfysize=6.5cm
\epsfxsize=7.5cm
\centerline{\epsffile{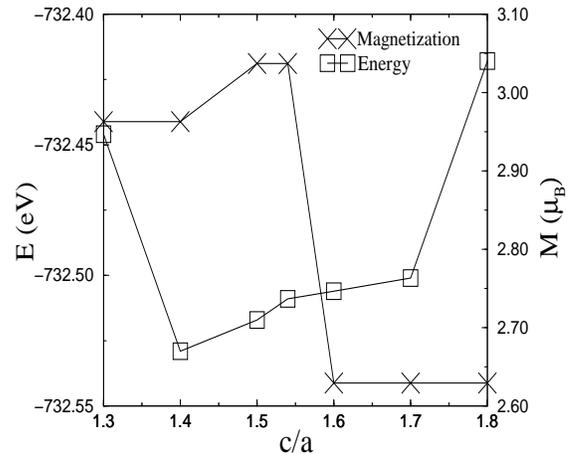}}
\caption{Total energy ($\Box $ on the left-hand side scale) and magnetization 
($\times$ on the right-hand side scale) as a function of the cell aspect ratio
c/a. For bulk MnAs c/a=1.54.}
\label{Fig5}
\end{figure}

We finally look at the magnetization and the total energy of distorted cells.
We consider cells with a volume equal to that of bulk MnAs and calculate the
energy and the magnetization as a function of the cell aspect ratio c/a. 
Our results are presented in Fig.\ref{Fig5}
Several interesting aspects must be pointed out. First note that the energy
scale is much smaller than that of Fig.\ref{Fig3}. The total energy for
aspect ratios from 1.4 to 1.7 changes only by 0.25~meV (which is equal to $k_BT$
at room temperature). By comparing figures \ref{Fig3} and
\ref{Fig5} one can conclude that NiAs-type MnAs can accommodate large distortions,
which are much less energy demanding than volume changes.  
This can be easily understood by using a simple geometrical argument.
Consider the nearest neighbor distance d$_{\mathrm Mn-As}$ as a function of
the lattice constant a for fixed cell volume
\begin{equation}
{\mathrm{d}_{\mathrm Mn-As}}=\frac{\mathrm a}{\sqrt{3}}\sqrt{1+\frac{V_0^2}{\mathrm a^6}},
\end{equation}
where $V_0$ is the volume occupied by a Mn-As pair. It clear that 
d$_{\mathrm Mn-As}$ is non-monotonic with a minimum at 
${\mathrm a}=(2V_0)^{1/6}$, which is 3.629\AA\ in the case of bulk MnAs. Moreover 
d$_{\mathrm Mn-As}$ does not vary much with the lattice constant a around
this minimum. For c/a increasing from 1.3 to 1.8, a
increases by about 10\% while d$_{\mathrm Mn-As}$ increases by $\le1$\%.
Since the total energy does not vary considerably as a
function of the Mn-As bond angle but only as a function of its bond length, 
we can conclude that the Mn-As bond is largely ionic and
dominates the total energy. 
In fact for the lattice distortions considered the second nearest neighbor 
distances (As-As and Mn-Mn) change drastically, while the first nearest neighbor
distance (Mn-As) is almost constant. This produces no appreciable change in
the total energy suggesting that the latter is dominated only by the Mn-As bond.
Moreover because of the absence of any strong angular dependence we can also 
conclude that the Mn-As has a large ionic component.

Our calculated tiny energy variation for even large distortions, combined with the large 
change in magnetization at small distortions explain most of the properties 
of the different epitaxial orientations of MnAs
onto GaAs. In fact it is worth noting that a large change of the
magnetization occurs for ratios c/a just above the one corresponding to the 
bulk. The magnetization passes from a high magnetization region c/a$<$1.54 to a
low magnetization region c/a$>$1.54. This is very important since large changes
of the magnetization have been observed in MnAs grown onto GaAs with different
crystalline orientations (type A, type B and MnAs onto (111) GaAs).
It is difficult to correlate directly the deformation of the cell with the
magnetization observed in experiments
since a complicated surface reconstruction occurs \cite{Schi99}
and a detailed characterization is not easy. We do not want to go into details
of this intricate issue and we only suggest that the reduction of the
magnetization of MnAs grown onto GaAs may be due to large deformation of the
hexagonal lattice.

\section{Zincblende MnAs}

\begin{figure}[htbp]
\narrowtext
\epsfysize=6cm
\epsfxsize=8cm
\centerline{\epsffile{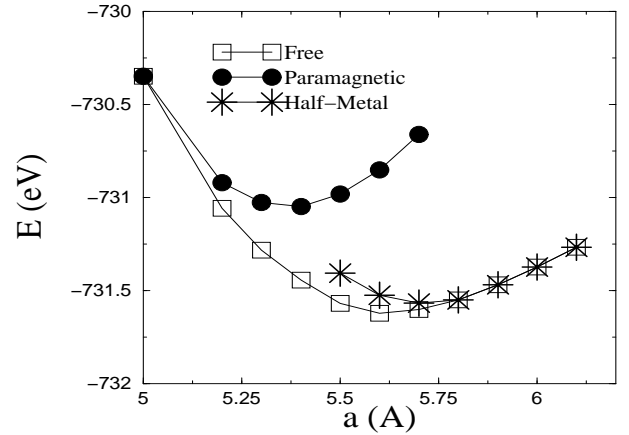}}
\caption{Total energy as a function of the zincblende lattice spacing for the
paramagnetic phase ($\bullet$), the half-metallic phase ($*$) and for
non-constrained magnetic
moment ($\Box$).}
\label{Fig6}
\end{figure}

\begin{figure}[htbp]
\narrowtext
\epsfysize=7cm
\epsfxsize=8cm
\centerline{\epsffile{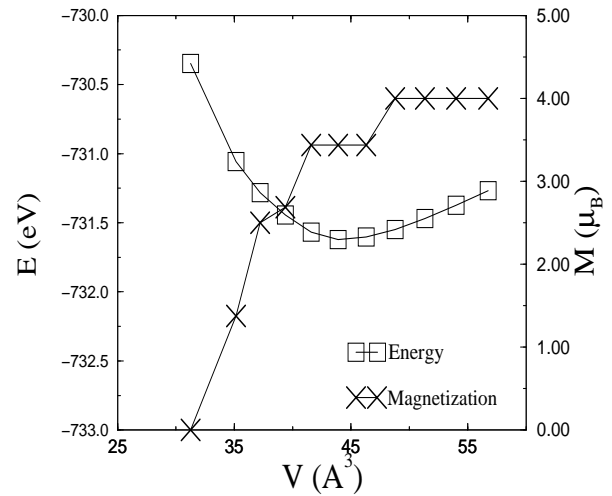}}
\caption{Total energy ($\Box $ on the left-hand side scale) and magnetization 
($\times$ on the right-hand side scale) as a function the unit cell volume.
The half-metallic behavior occurs for ${\mathrm a}\ge 5.8$\AA.}
\label{Fig7}
\end{figure}

\begin{figure}[htbp]
\narrowtext
\epsfysize=7.7cm
\epsfxsize=9cm
\centerline{\epsffile{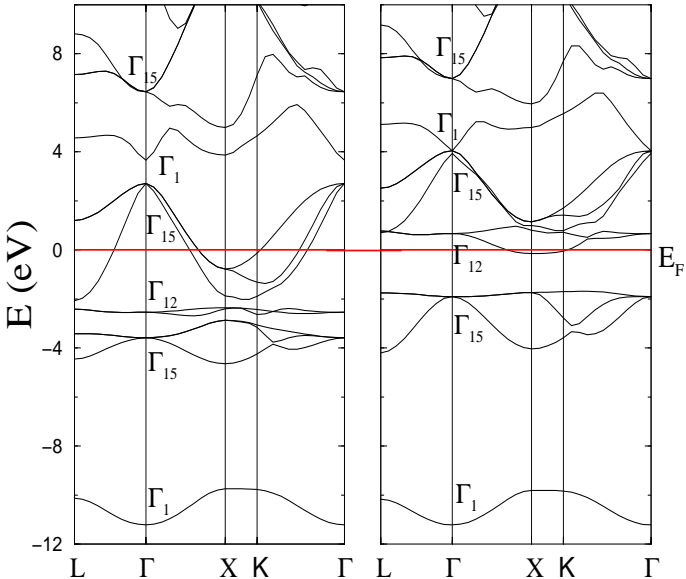}}
\caption{Band structure for zincblende MnAs at the LDA energy minimum (a=5.7
\AA ).
The figure on the left corresponds to the majority spin and the one on the 
right to the minority. The horizontal line denotes the position of the Fermi 
energy, which has been chosen to be 0~eV. }
\label{Fig8}
\end{figure}

We now consider MnAs with the zincblende lattice structure. 
In Fig.\ref{Fig6} we present
the total energy as a function of the lattice constant for MnAs and
also compare the total energy of the same structure when the magnetic
moment is fixed either to 0$\mu_B$ (paramagnetic phase) and to 
4$\mu_B$ (half-metallic phase).
The equilibrium lattice constant is found between 5.6\AA\ and
5.7\AA\ which is very similar to the lattice
constant of GaAs. Such a value is smaller (as we would expect within the LDA) 
than that predicted from the linear extrapolation of the experimental
lattice constant of Ga$_{1-x}$Mn$_x$As 
for $x\rightarrow1$ (5.89\AA) \cite{Ohno99}. It is also smaller than another LDA
result obtained with the FLAPW method (5.9\AA) \cite{Masa98}.
This could be due to the pseudopotential method used here or
differences in convergence or sampling between the two calculations.
We also notice that the paramagnetic phase becomes energetically favorable for
a=5.0\AA , where a ferromagnetic to paramagnetic transition is predicted.
In contrast, for
lattice spacings larger than 5.8\AA\ a transition to an half-metallic state 
is predicted. This is a very important result since half-metallicity is one of the
conditions required
to obtain large signals in a semiconductor spin-valve \cite{Schm99}.
The behavior of the magnetization as a function of the unit cell volume is
summarized in Fig.\ref{Fig7}.
We notice that Fig.\ref{Fig7} looks similar to its counterpart for the NiAs-type
structure (Fig.\ref{Fig3}) although in the latter no transition to an 
half-metallic state is found. 
This fundamental difference between the two structures can
be understood by looking at the band structure of zincblende MnAs in
Fig.\ref{Fig8}.

We first note that, excluding the presence of the Mn d bands, the bandstructure
closely resembles that of the non-magnetic III-V semiconductors. Consider the
majority bands first. By symmetry analysis
we can easily identify the lower lying As s states (lowest
band, $\Gamma_1$ at the $\Gamma$ point), the As p valence band (first
$\Gamma_{15}$ point above $E_{\mathrm F}$) and the first of the conduction 
bands (first $\Gamma_1$ point above $E_{\mathrm F}$). However, due to the strong
interaction with the Mn d states, the As p bands 
(top of the valence bands in GaAs) are pushed toward higher
energies and become half filled. The Mn d bands, which are split into the
doubly degenerate $e_g$ band ($\Gamma_{12}$) and the triply degenerate $t_{2g}$
band ($\Gamma_{15}$) are below the Fermi energy and entirely occupied. 
Therefore one can conclude that two important interactions
are present: i) the s-p interaction giving rise to the $sp^3$ bond typical of
non-magnetic semiconductors and ii) the p-d interaction responsible for the
magnetism, which sets the position and the dispersion of the As p band. Note also
that in contrast to the prediction from crystal field theory, the $e_g$
orbitals possess higher energy than the $t_{2g}$. This is also due to the strong
p-d interaction.
If we now turn our attention to the minority band, we can find most of the
features of the majority. The main difference is the large split between the 
$t_{2g}$ and the $e_g$ states, which gives rise to a large gap in the band
structure. 
We observe that the Fermi energy cuts through the band center of the high
dispersion p band in the majority band, and through the band edge of the almost
dispersionless d band in the minority. This is very important from the point of
view of the transport, since very different spin-dependent effective masses are
predicted.
When the lattice constant is increased, we expect a general narrowing
of all the bands. This does not affect qualitatively the majority band. 
However for some critical lattice spacing the band edge of the $e_g$ band in
the minority spin shifts above $E_{\mathrm F}$ giving rise to a semiconducting
behavior. This generates the half metallicity.

In summary, if we start from the atomic configuration for Mn and As, the
mechanism giving rise to the magnetism is the transfer of one electron from the 
Mn d orbitals to the As p shell in order to form the $sp^3$ bond. 
The remaining 4 electrons in the Mn d shell
maximize the local magnetic moment of Mn due to Hund's coupling and the final
state turns out to be ferromagnetic due to the strong p-d interaction. Note
that these are features of the tetrahedral coordination and are absent in the
NiAs-type structure, where no half metallic state is predicted.

\begin{figure}[htbp]
\narrowtext
\epsfysize=9.7cm
\epsfxsize=8cm
\centerline{\epsffile{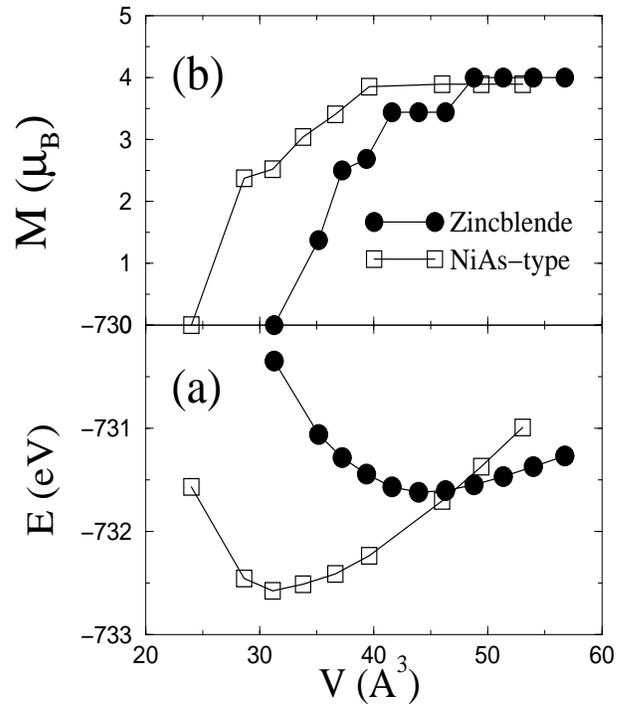}}
\caption{Total energy (a) and magnetization (b) as a function of the MnAs pair
volume for the zincblende ($\bullet$) and the NiAs-type structure
($\Box$). Note the large stability of the NiAs-type structure over a very broad
volume range.}
\label{Fig9}
\end{figure}

Finally we look at the stability of the zincblende structure with respect to the
NiAs-type. The total energy and the saturation magnetization of both the
structures as a function of the volume occupied by a MnAs pair are presented in
Fig.\ref{Fig9}.
It is clear that the NiAs-type has a much lower total energy and also a more
compressed lattice. Therefore it is the stable structure at all the
thermodynamically accessible pressures. This is consistent with the very small
dilution limit of Mn in zincblende GaAs and with the fact that the annealing usually
induces segregation of MnAs NiAs-type particles within the GaAs matrix 
\cite{Ohno99,Ohno98}. 

We also note that for very expanded unit cells the
zincblende structure becomes stable with a crossover volume which corresponds
to a zincblende lattice constant of 5.8\AA. However this large volume
increase in the NiAs-type structure is very unlikely to occur, even if the lattice
constants are forced to be large by growing onto substrates with large mismatch.
As we noted previously, the NiAs-type structure can easily accommodate large
cell distortions (Fig.\ref{Fig5}). Therefore, at equilibrium it is energetically
more favorable for the system to distort the cell, instead of increasing the volume
and inducing a NiAs-type to zincblende transition. 

\begin{figure}[htbp]
\narrowtext
\epsfysize=7.7cm
\epsfxsize=8.5cm
\centerline{\epsffile{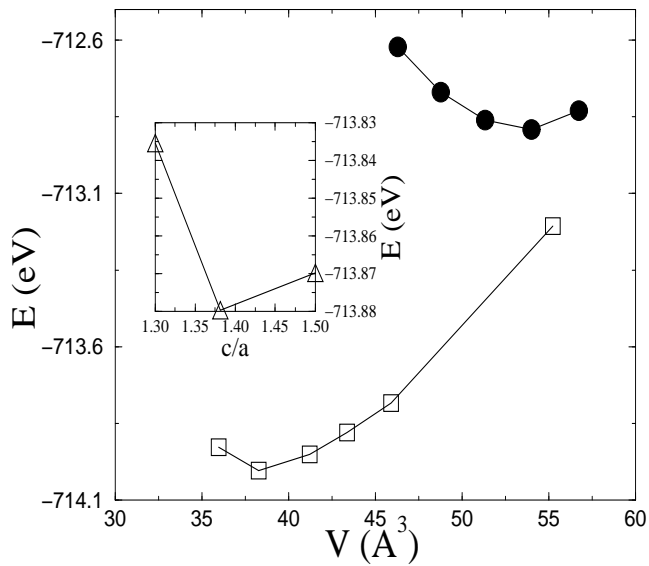}}
\caption{Total energy as a function of the MnBi pair
volume for the zincblende ($\bullet$) and the NiAs-type structure
($\Box$). Note the large stability of the NiAs-type structure over a very broad
volume range as in the MnAs case. In the inset the energies for three 
different values of c/a for a volume corresponding to the bulk MnBi are shown. 
The central point corresponds to bulk MnBi ($V=43.37$\AA$^3$).}
\label{Fig10}
\end{figure}

Finally we investigate the effect of the anion size on the stability properties
of the zincblende structure. Our hope is
that for larger ionic radii the NiAs-type structure should become more rigid and a
transition to the zincblende structure at large volume may be induced. 
We perform the same calculations described above for MnBi both with
NiAs-type and zincblende structure (bulk MnBi is a ferromagnetic metal at room
temperature with NiAs-type structure and lattice constants a=4.17\AA, 
c=5.764\AA). We did not find any relevant qualitative difference with respect 
to the MnAs case (see Fig.\ref{Fig10}) and this is related to the geometrical 
argument presented in section IV.
Hence we conclude that, for all the Mn-pnictides,
the NiAs-type is the more stable crystalline structure and a transition to
zincblende structure can be obtained only with highly non-equilibrium methods.

\section{Mn-As bond in the zincblende structure}

Finally, to analyze the detailed nature of the
Mn-As bonding in the zincblende lattice 
we perform a tight-binding fit to our calculated LDA band structure. 
We use a fitting algorithm
included in the package OXON \cite{OXON} which minimizes the following function
\begin{equation}
f(E_n,\vec{\gamma})=\sum_n\alpha_n|E_n(k)-E^c_n(k,\vec{\gamma})|,
\end{equation}
where $\vec{\gamma}$ is the $m$-dimensional vector containing the Slater-Koster
tight-binding parameters \cite{Slat54}, $\alpha_n$ is the weight assigned to the
eigenvalue $E_n$ and $E_n^c$ is the computed eigenvalue.
We consider first (Mn-As) and second (Mn-Mn and As-As) nearest neighbor
couplings. This interaction set has been shown to reproduce correctly the valence band 
and the first of the conduction bands of diamond-like semiconductors 
\cite{Chad75}. In the case of MnAs the presence of the Mn d band makes the
fitting procedure more complicated then in non-magnetic semiconductors. 
However we do not need to introduce any {\it ad hoc} excited state (such as $s^*$) to fit
the valence band and the first conduction band
as is frequently done in the literature \cite{Vogl83}. 
The fit has been performed using 350 eigenvalues which correspond to the lowest
10 energies calculated at 35 $k$-vectors and the weight is chosen to be 1 for
each eigenvalue. We perform two independent fits for the majority and minority
band allowing both the on-site energies and the hopping integrals to be
different. 

\begin{figure}[htbp]
\narrowtext
\epsfysize=7.7cm
\epsfxsize=8cm
\centerline{\epsffile{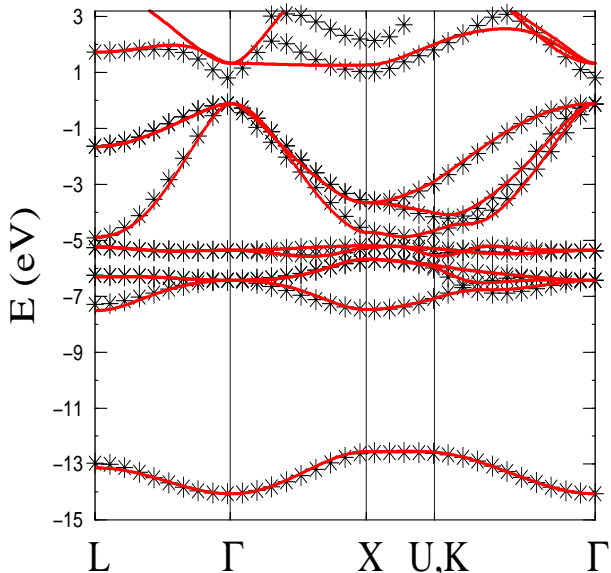}}
\caption{Fit of the band structure for the majority band of MnAs. The stars ($*$)
are the {\it ab initio} eigenvalues and the continuous line are the band calculated
with the parameters of Tables \ref{Tab1} and \ref{Tab2}.}
\label{Fig11}
\end{figure}

\begin{figure}[htbp]
\narrowtext
\epsfysize=7.7cm
\epsfxsize=8cm
\centerline{\epsffile{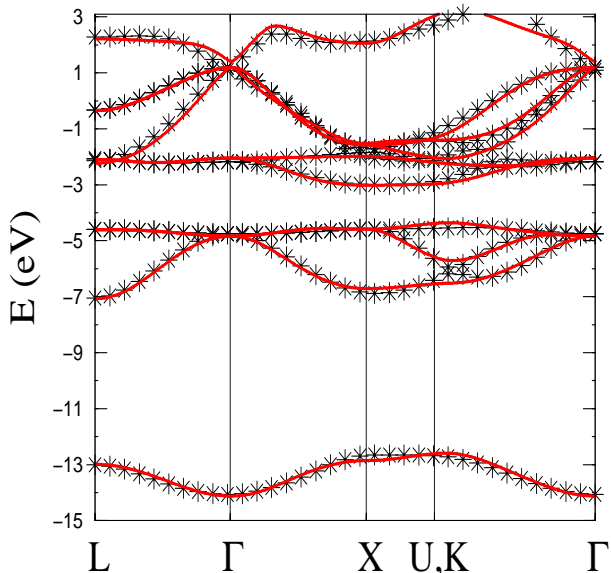}}
\caption{Fit of the band structure for the minority band of MnAs. The stars ($*$)
are the {\it ab initio} eigenvalues and the continuous line are the band calculated
with the parameters of Tables \ref{Tab1} and \ref{Tab2}.}
\label{Fig12}
\end{figure}

In figures \ref{Fig11} and \ref{Fig12} we present the best fits for the
majority and minority spin band respectively. 
We note that the agreement with the {\it ab initio}
bands is remarkably good, in particular for the lower lying bands. 
The highest bands are not well reproduced
(in particular for the majority spin) because we did not include
higher energy bands in the fit. The complete parameterization giving rise 
to the bands in figures \ref{Fig11} and \ref{Fig12} is provided in tables 
\ref{Tab1} and \ref{Tab2}.

Several important points must be stressed. First we note that the major
difference between the majority and the minority parameters is the value of
the Mn 3d and As 4p on-site energies. The on-site energy of the 3d Mn orbitals
coincides with the position of the $e_g$ states, which by symmetry
are weakly coupled to the
other bands. By contrast the $t_{2g}$ states are strongly coupled to the
above As p band and this coupling is strongly spin dependent. From table \ref{Tab2}
one can see that the dominant hopping integrals are along the Mn-As bond and
that the parameters are quite similar for the two spin directions. The only 
parameters which differ strongly for up and down spin are
the sd$\sigma$, pd$\sigma$ and pd$\pi$ elements. These govern the repulsion of the
$t_{2g}$ Mn d bands and the As p bands and are responsible for the large
$e_g$-$t_{2g}$ splitting in the minority band.
Finally we check our hypothesis on the formation of a strong $sp^3$ bond by
comparing the ss, sp and pp hopping integrals for MnAs with those of GaAs, which
have been obtained with the same fitting procedure from our calculated LDA band
structure.
The GaAs parameters are shown in table \ref{Tab3} and their magnitude is clearly
comparable with the MnAs case. Moreover also the second nearest neighbor 
pp$\sigma$ and pp$\pi$ integrals, which are the only 
second nearest neighbor parameters with appreciable amplitude 
and the ones that 
describe most of the dispersion of the
As p band along K$\rightarrow\Gamma$, are very similar.
This clearly suggests the formation of a strong $sp^3$ bond in MnAs together 
with a strong p-d hybridation. 

\section{Conclusion}

We have investigated theoretically the structural and magnetic properties of 
the hexagonal NiAs-type and the zincblende phases of 
MnAs. From the analysis we
conclude that the NiAs-type structure is more stable at all thermodynamic
pressures. A phase transition to the zincblende structure is predicted for
volume stretching, which in principle could be obtained by growing onto a substrate
with large lattice mismatch.
However the NiAs phase can accommodate large lattice distortions which
preserve the Mn-As nearest neighbor distance and a transition to a zincblende
phase looks very unlikely.

We have also investigated the zincblende phase in more detail, since the 
tetrahedral coordination is crucial for the physics of diluted magnetic 
semiconductors. By performing
both {\it ab initio} calculations and tight-binding fitting we see that two
important effects are present: i) the formation of a stable $sp^3$ bond and ii) a
strong p-d hybridation, which splits the Mn d bands and gives rise to
ferromagnetism. In particular for large lattice spacings the zincblende structure
is predicted to be half metallic.

\vspace{0.3in}

{\bf Acknowledgments}: 
We would like to thank G.~Theurich for having provided and supported the use 
SPECTER and M.~Fearn and A.P.~Horsfield for the tight-binding package OXON.
This work made use of MRL Central Facilities supported by the National Science 
Foundation under award No. DMR96-32716.
This work is supported by the DARPA/ONR under the grant N0014-99-1-1096,
by ONR grant N00014-00-10557 and by
NSF-DMR under the grant 9973076.
Useful discussions with N.~Samarth, D.D.~Awschalom and L.~Sham are kindly 
acknowledged.

\end{multicols}

\newpage

\begin{center}
\begin{table}
\begin{tabular}{ccc}
  &  \  \ \ \ Majority (eV) \ \ \ & \ \ \ Minority (eV) \ \ \ \\
\hline
E$_{\mathrm{As}(4s)}$ & -9.8394 & -9.9188 \\
E$_{\mathrm{As}(4p)}$ & -0.8709 & -0.3476 \\
E$_{\mathrm{Mn}(4s)}$ &  0.7380 & -1.0519 \\
E$_{\mathrm{Mn}(4p)}$ &  1.1254 &  1.4601 \\
E$_{\mathrm{Mn}(3d)}$ & -5.4699 & -2.3417 \\ \hline
\end{tabular}
\caption{On site energies for the majority and minority bands of zincblende MnAs. 
The notation is that of Slater and Koster [37].}
\label{Tab1}
\end{table}
\end{center}
\begin{center}
\begin{table}
\begin{tabular}{ccc}
  &  \ \ \ Majority (eV) \ \ \ & \ \ \ Minority (eV) \ \ \ \\
\hline
E$_{[\mathrm{As}(4s)-\mathrm{Mn}(4s)]\sigma}$ &  1.8554 &  1.3592 \\
E$_{[\mathrm{As}(4s)-\mathrm{Mn}(4p)]\sigma}$ & -1.8684 & -1.5061 \\
E$_{[\mathrm{As}(4s)-\mathrm{Mn}(3d)]\sigma}$ &  1.0852 &  2.0521 \\
E$_{[\mathrm{As}(4p)-\mathrm{Mn}(4s)]\sigma}$ &  2.5898 &  2.3806 \\
E$_{[\mathrm{As}(4p)-\mathrm{Mn}(4p)]\sigma}$ &  2.2825 &  2.5146 \\
E$_{[\mathrm{As}(4p)-\mathrm{Mn}(4p)]\pi}$ &    -0.9342 & -0.8346 \\
E$_{[\mathrm{As}(4p)-\mathrm{Mn}(3d)]\sigma}$ &  1.0329 &  1.4592 \\
E$_{[\mathrm{As}(4p)-\mathrm{Mn}(3d)]\pi}$ &    -0.4317 & -0.6831 \\ \hline
E$_{[\mathrm{As}(4s)-\mathrm{As}(4s)]\sigma}$ & -0.0133 & -0.1377 \\
E$_{[\mathrm{As}(4s)-\mathrm{As}(4p)]\sigma}$ &  0.0031 &  0.0175 \\
E$_{[\mathrm{As}(4p)-\mathrm{As}(4p)]\sigma}$ &  0.0009 &  0.0064 \\
E$_{[\mathrm{As}(4p)-\mathrm{As}(4p)]\pi}$ &     0.0255 & -0.0373 \\ \hline
E$_{[\mathrm{Mn}(4s)-\mathrm{Mn}(4s)]\sigma}$ &  0.0000 &  0.0000 \\
E$_{[\mathrm{Mn}(4s)-\mathrm{Mn}(4p)]\sigma}$ &  0.0053 & -0.0009 \\
E$_{[\mathrm{Mn}(4s)-\mathrm{Mn}(3d)]\sigma}$ & -0.1579 &  0.1216 \\
E$_{[\mathrm{Mn}(4p)-\mathrm{Mn}(4p)]\sigma}$ &  0.4411 &  0.3323 \\
E$_{[\mathrm{Mn}(4p)-\mathrm{Mn}(4p)]\pi}$ &    -0.2265 & -0.0167 \\
E$_{[\mathrm{Mn}(4p)-\mathrm{Mn}(3d)]\sigma}$ &  0.2398 & -0.2232 \\
E$_{[\mathrm{Mn}(4p)-\mathrm{Mn}(3d)]\pi}$ &    -0.0298 &  0.1963 \\
E$_{[\mathrm{Mn}(3d)-\mathrm{Mn}(3d)]\sigma}$ & -0.1139 & -0.1199 \\
E$_{[\mathrm{Mn}(3d)-\mathrm{Mn}(3d)]\pi}$ &     0.0453 &  0.0773 \\
E$_{[\mathrm{Mn}(3d)-\mathrm{Mn}(3d)]\delta}$ & -0.0002 & -0.0005 \\  \hline
\end{tabular}
\caption{Hopping integrals for the majority and minority bands of zincblende
MnAs. The notation is that of Slater and Koster [37]. } 
\label{Tab2}
\end{table}
\end{center}
\begin{center}
\begin{table}
\begin{tabular}{cc}
  &  \ \ \ GaAs (eV) \ \ \ \\
\hline
E$_{\mathrm{As}(4s)}$ & -7.7859  \\
E$_{\mathrm{As}(4p)}$ &  1.0395  \\
E$_{\mathrm{Ga}(4s)}$ & -0.2676  \\
E$_{\mathrm{Ga}(4p)}$ &  2.8029  \\ \hline
E$_{[\mathrm{As}(4s)-\mathrm{Ga}(4s)]\sigma}$ &  1.4895 \\
E$_{[\mathrm{As}(4s)-\mathrm{Ga}(4p)]\sigma}$ & -1.5445 \\
E$_{[\mathrm{As}(4p)-\mathrm{Ga}(4s)]\sigma}$ &  2.5864 \\
E$_{[\mathrm{As}(4p)-\mathrm{Ga}(4p)]\sigma}$ &  2.1571 \\
E$_{[\mathrm{As}(4p)-\mathrm{Ga}(4p)]\pi}$    & -0.7540 \\ \hline
E$_{[\mathrm{As}(4s)-\mathrm{As}(4s)]\sigma}$ &  0.0032 \\
E$_{[\mathrm{As}(4s)-\mathrm{As}(4p)]\sigma}$ & -0.0437 \\
E$_{[\mathrm{As}(4p)-\mathrm{As}(4p)]\sigma}$ & -0.0009 \\
E$_{[\mathrm{As}(4p)-\mathrm{As}(4p)]\pi}$ &     0.0010 \\ \hline
E$_{[\mathrm{Ga}(4s)-\mathrm{Ga}(4s)]\sigma}$ & -0.1038 \\
E$_{[\mathrm{Ga}(4s)-\mathrm{Ga}(4p)]\sigma}$ & -0.0919 \\
E$_{[\mathrm{Ga}(4p)-\mathrm{Ga}(4p)]\sigma}$ &  0.4828 \\
E$_{[\mathrm{Ga}(4p)-\mathrm{Ga}(4p)]\pi}$ &    -0.1512 \\ \hline
\end{tabular}
\caption{Tight-binding parameters for GaAs. The notation
is that of Slater and Koster [37]. }
\label{Tab3}
\end{table}
\end{center}


\begin{references}
\bibitem{Pri95} G.~Prinz, Phys. Today {\bf 48}, 58 (1995) 

\bibitem{Aws1} J.M.~Kikkawa and D.D.~Awschalom, Phys. Rev. Lett. {\bf 80}, 4113
(1998) 

\bibitem{Aws2} J.M.~Kikkawa and D.D.~Awschalom, Nature {\bf 397}, 139 (1998)

\bibitem{QC} D.P.~Di Vincenzo, Science {\bf 270}, 255 (1995)

\bibitem{Ohno99} H.~Ohno, J. Magn. Magn. Matter {\bf 200}, 110 (1999) 

\bibitem{Ohno98} H.~Ohno, Science {\bf 281}, 951 (1998) 

\bibitem{Rouk00} H.X.~Tang, F.G.~Monzon, R.~Lifshitz, M.C.~Cross and
M.L.~Roukes, 
Phys. Rev. B {\bf 61}, 4437 (2000) and reference therein

\bibitem{Lee99} W.Y.~Lee, S.~Gardelis, B.C.~Choi, Y.B.~Xu, C.G.~Schmidt,
C.H.W.~Barnes, D.A.~Ritchie, E.H.~Linfield and J.A.C.~Bland, Appl. Phys. Lett.
{\bf 85}, 6682 (1999) 

\bibitem{Ham99}P.R.~Hammer, B.R.~Bennet, M.J.~Yang and M.~Johnson, Phys. Rev. Lett.
{\bf 83}, 203 (1999) 

\bibitem{Awch99} Y.~Ohno, D.K.~Young, B.~Beschoten, F.~Matsukura and H.~Ohno,
D.D.~Awschalom, Nature {\bf 402}, 790 (1999)

\bibitem{Fied99} R.~Fiederling, M.~Keim, G.~Reuscher, W.~Ossau, G.~Schmidt,
A.~Waag and L.W.~Molenkamp, Nature {\bf 402}, 787 (1999)

\bibitem{Schm99} G.~Schmidt, L.W.~Molenkamp, A.T.~Filip and B.J.~van~Wees,
cond-mat/9911014

\bibitem{Tan99-1} M.~Tanaka, K.~Saito and T.~Nishinaga, Appl. Phys. Lett. {\bf
74}, 64 (1999) and reference therein

\bibitem{Tan93} M.~Tanaka, J.P.~Harbison, T.~Sands, B.~Philips, T.L.~Cheeks,
J.~De~Boeck, L.T.~Florez and V.G.~Keramidas, Appl. Phys. Lett. {\bf 63}, 696 
(1993)
%

\bibitem{Tan94-1} M.~Tanaka, J.P.~Harbison, T.~Sands, T.L.~Cheeks and 
V.G.~Keramidas, J. Vac. Technol. B {\bf 12}, 1091 (1994) 
%

\bibitem{Ake95} K.~Akeura, M.~Tanaka, M.~Ueki and T.~Nishinaga, Appl. Phys.
Lett. {\bf 67}, 3349 (1995)
%

\bibitem{Oga99} T.~Ogawa, M.~Shirai, N.~Suzuki and I.~Kitagawa, J. Magn. Magn.
Mater. {\bf 196-197}, 428 (1999)

\bibitem{Kohn64} H.~Hohenberg and W.~Kohn, Phys. Rev. {\bf 136}, B864 (1964),
W.~Kohn and L.~Sham, Phys. Rev. {\bf 140} A1133 (1965) 

\bibitem{Nicola99} N.A.~Hill and K.M.~Rabe, Phys. Rev. B {\bf 59}, 8759 (1999)

\bibitem{Rap90} A.M.~Rappe, K.M.~Rabe, E.~Kaxiras and J.D.~Ioannopolous, Phys.
Rev. B {\bf 41}, 1227 (1990)

\bibitem{Kle82} L.~Kleinman and D.M.~Bylander, Phys. Rev. B {\bf 48}, 1425
(1982)

\bibitem{Blo90} P.E.~Bl\"ochl, Phys. Rev. B {\bf 41}, 5414 (1990)

\bibitem{Ham79} D.R.~Hamann, M.~Sch\"ulter and C.~Chiang, Phys. Rev. Lett. {\bf
43}, 1494 (1979)

\bibitem{Ger00} G.~Theurich, B.~Anson, N.A.~Hill and A.~Hill, preprint submitted to
Computing in Science and Engineering. The code is pubblically available at 
http://www.mrl.ucsb.edu/$\sim$theurich/Gespenst/

\bibitem{Payne92} M.C.~Payne, M.P.~Teter, D.C.~Allan, T.A.~Arias and
J.D.~Joannopoulos, Rev. Mod. Phys. {\bf 64}, 1045 (1992)

\bibitem{Kerk81} G.P.~Kerker, Phys. Rev. B {\bf 23}, 3082 (1981)
%

\bibitem{Moti86} K.~Motizuki, K.~Katoh and A.~Yanase, J. Phys. C: Solid State
Phys. {\bf 19}, 495 (1986) and references therein

\bibitem{Tan94} M.~Tanaka, J.P.~Harbison, M.C.~Park, Y.S.~Park, T.~Shin and
G.M.~Rothberg, Appl. Phys. Lett. {\bf 65}, 1964 (1994)

\bibitem{Tan99} M.~Tanaka, K.~Saito, M.~Goto and T.~Nishinaga, J. Magn. Magn.
Matter. {\bf 198-199}, 719 (1999)

\bibitem{Tan94b} M.~Tanaka, J.P.~Harbison, M.C.~Park, Y.S.~Park, T.~Shin and
G.M.~Rothberg, J. Appl. Phys. {\bf 76}, 6278 (1994)

\bibitem{Sam} N.~Samarth, private communication 

\bibitem{Maki98} K.~Maki, T.~Kaneko, H.~Hiroyoshi and K.~Kamigaki, J. Magn. Magn.
Mater. {\bf 177-181}, 1361 (1998)

\bibitem{Yam83} Y.~Yamaguchi and H.~Watanabe, J. Magn. Magn. Mater. {\bf 31-34},
619 (1983)

\bibitem{Schi99} F.~Schippan, A.~Trampert, L.~D\"aweritz and K.H.~Ploog,
J. Vac. Sci. Technol. B {\bf 17}, 1716 (1999)

\bibitem{Masa98} M.~Shirai, T.~Ogawa, I.~Kitagawa and N.~Suzuki, J. Magn. Magn.
Mater. {\bf 177-181}, 1383 (1998)

\bibitem{OXON} OXON (Oxford O(N) tight binding
code) was developed at The Materials Modelling Laboratory of the
Department of Materials at the University of Oxford.

\bibitem{Slat54} J.C.~Slater and G.F.~Koster, Phys. Rev. {\bf 94}, 1498 (1954)

\bibitem{Chad75} D.J.~Chadi and M.L.~Cohen, Phys. Stat. Sol. {\bf 68}, 405 (1975)

\bibitem{Vogl83} P.~Vogl, H.P.~Hjalmarson and J.D.~Dow, J. Phys. Chem. Solids
{\bf 44}, 365 (1983)

\end{references}
\end{document}